\documentclass[a4paper,showkeys,floatfix,aps,pre,longbibliography,superscriptaddress,reprint]{revtex4-1}
\usepackage{graphics,graphicx}
\usepackage{amsmath,amssymb}

\usepackage{graphics,graphicx}
\usepackage{dcolumn,bm}
\usepackage{psfrag}
\usepackage{xstring}
\usepackage{color}
\newcommand{\srm}
{\affiliation{Department of Physics, SRM University - AP, Amaravati,
 Andhra Pradesh - 522240, India}}

\newcommand{\srmcse}
{\affiliation{Department of Computer Science and Engineering, SRM University - AP, Amaravati,
 Andhra Pradesh - 522240, India}}

 \newcommand{\saha}
 {\affiliation{Condensed Matter Physics Division, Saha Institute of Nuclear Physics, Kolkata
700064, India}}

\begin{document}

\title{Quantum Annealing in SK Model Employing Suzuki-Kubo-deGennes
Quantum Ising Mean Field Dynamics}

\author{Soumyaditya Das}

\email{soumyaditya\_das@srmap.edu.in}
\srm
\author{Soumyajyoti Biswas}

\email{soumyajyoti.b@srmap.edu.in}
\srm
\srmcse
\author{Bikas K. Chakrabarti}
\email{bikask.chakrabarti@saha.ac.in}
\saha

\begin{abstract}
We study a quantum annealing approach for estimating the ground state energy of the Sherrington-Kirpatrick mean field spin glass model using the Suzuki-Kubo-deGennes dynamics applied for individual local magnetization components. The solutions of the coupled differential equations, in discretized state, give a fast annealing algorithm (cost $N^3$) in estimating the ground state of the model: Classical ($E^0= -0.7629 \pm 0.0002$), Quantum ($E^0=-0.7623 \pm 0.0001$) and Mixed ($E^0=-0.7626 \pm 0.0001$), all of which are to be compared with the best known estimate $E^0= -0.763166726 \dots$ . We infer that the continuous nature of the magnetization variable used in the dynamics here is the reason for reaching close to the ground state quickly and also the reason for not observing the de-Almeida-Thouless line in this approach.
\end{abstract}
\maketitle

\section{Introduction}
The Sherrington-Kirkpatrick (SK) model
\cite{SK1975} was proposed as a mean-field or
long-range Ising model where the spin-spin
interactions are random, causing frustra-
tions (see e.g. \cite{BinderYoung1986} for a review). It is
well-known, because of the random
frustrating constraints in the system,
the search of the (degenerate) ground
state and estimating the corresponding
energy are nontrivial \cite{BinderYoung1986,ChakrabartiPhilTrans2022}. Indeed, the
rugged (free) energy landscape was
already characterized by occasional
barrier heights of the order of system
size, disallowing the discrete Ising
spin dynamics to anneal the system to
its ground state(s) \cite{ray89,ChakrabartiPhilTrans2022,DasChakrabarti2008,RajakPhilTrans2022}.

The continuous mean field dynamics of
thermal average of the local magnetization
of individual Ising spins, formulated by
Suzuki and Kubo \cite{SuzukiKubo1968} in 1968, has already
been shown \cite{DasPRE2025} to help annealing (from the
paramagnetic phase above the glass
transition temperature) the SK spin glass
smoothly very close to the ground state,
when augmented by a modified \cite{DasPRE2025} Thouless-
Anderson-Palmer (TAP) reaction field \cite{BinderYoung1986,TAP1977}. It can lead  quickly \cite{DasPRE2025} to some
states close to the ground state of the
SK model (for system size upto 10,000 spins) with energy per spin to be
around $0.7629 \pm 0.0002$ (compared to
the best available estimate \cite{Oppermann2007} 0.7631667265... for
the full Replica Symmetry Breaking or
RSB solution \cite{Parisi1980}). We will augment this
Suzuki-Kubo mean field dynamics with
the de Gennes \cite{deGennes1963} and Brout-Muller-Thomas
\cite{Brout1966} type mean field dynamics (following
\cite{SenChakrabarti1989,AcharyyaPhysicaA1994,AcharyyaJPhysA1994}) here for quantum annealing
of the SK model under transverse field and study quantum annealing of the SK model.

\section{ Mean field equation}
The Hamiltonian (for a particular
configuration [c] of the random exchange
interaction distribution) of  the SK model
in  presence of both transverse ($\Gamma$)
and longitudinal external magnetic
field ($h$) is represented as

\begin{equation}
 H^{[c]} = - \sum_{i,j} J_{ij}\sigma_i^z\sigma_j^z
-h \sum_i\sigma_i^z -\Gamma \sum_i \sigma_i^x. 
\end{equation}

\noindent Here, $\vec {\sigma}$ denotes the Pauli
spin vector, $h$ and $\Gamma$ are the external
longitudinal field and transverse field respectively,
$J_{ij}$ denotes  the long-range (randomly
ferromagnetic or antiferromagnetic) interaction
strength between the spins placed at $i$ -th and
$j$ -th sites in the SK model with Gaussian
distribution centered around zero:

\begin{equation}
P (J_{ij}) = (1/J)(N/2\pi)^{1/2}
\exp[-(N/2)(J_{ij}/J)^2],   \tag{1a}       
\end{equation}
\noindent with 
\begin{equation}
[J_{ij}^2]_{av} - [J_{ij}]_{av}^2
= J^2/N = 1/N. \tag{1b}
\end{equation}

It may be noted here that due to the
presence of transverse field (non-commuting
component of the cooperative part of the
Hamiltonian), the quantum dynamics of
$\sigma^z$ arises from the Heisenberg
equation of motion. However, one can
expect a simplified form of the dynamical
evolution in the mean-field approximation
(following refs. \cite{SuzukiKubo1968,DasPRE2025,TAP1977} and \cite{deGennes1963,Brout1966,SenChakrabarti1989,AcharyyaPhysicaA1994,AcharyyaJPhysA1994}).

The mean-field Hamiltonian can be written as

\begin{equation}
 H^{[c]}= - \sum_i  \vec{h}_i^{\text{eff}\,[c]}\cdot\vec{m_i}^{[c]}. 
\end{equation}

\noindent Here the effective mean-field
$\vec{h}_i^{\text{eff}\,[c]}$ on any spin
$\vec{m_i}$ for a particular
configuration [c] corresponding to a
particular realization of the
distribution of $J_{ij}$  has two
parts namely ${\vec{h}_i}^{z\,\text{eff}\,[c]}$,
coming from the standard Curie-Weiss
type cooperative interaction among
the spins (which is very small here
because of competing interactions) and
${\vec{h}_i}^{x\,\text{eff}\,[c]}$, the modified
(see \cite{DasPRE2025}) Thouless-Anderson-Palmer type
reaction field \cite{TAP1977}, coming from the
second order mean field effect:

\begin{equation}
   {\vec{h}_i^{\text{eff}\,[c]}}= {{h}^{z\,\text{eff}\,[c]}_i}\hat{z}+{h_i}^{x\,\text{eff}\,[c]}\hat{x},
\end{equation}
\noindent where
\begin{equation}
\left|\vec{h}_i^{\text{eff}\,[c]}\right| = \left[ \left( \sum_j J_{ij} m_j^{z\,[c]} +h -\left(1 - q^{[c]}\right)\, m_i^{z\,[c]} \right)^2 + \Gamma^2 \right]^{1/2}  \tag{4a}
\end{equation}
\noindent with 
\begin{equation}
   h_i^{z\,\text{eff}\,[c]}= \sum_j J_{ij} m_j^{z\,[c]} + h - \left(1 - q^{[c]}\right)\, m_i^{z\,[c]} \tag{4b}
\end{equation}
\noindent and 
\begin{equation}
h_i^{x\,\text{eff}\,[c]} = \Gamma. \tag{4c}
\end{equation}

\noindent Here $\vec{m}_i^{[c]} \equiv \langle\vec{\sigma}_i^{[c]}\rangle$, where $< \cdot >$
denotes the thermal average, and the spin
glass order parameter for a particular configuration $[c]$ is contributed by $<m^z>$ only, and is given by
\begin{equation}
q^{[c]} = \frac{1}{N} \sum_{i=1}^N \left[ \langle m_i^{z\,[c]} \rangle^2 \,  \right], \hskip0.5cm  q=\overline{q^{[c]}} \tag{5}
\end{equation} 
where the overhead bar denotes the average over the configurations $[c]$. Here in Eq. 4(b),  the reaction term is modified (as mentioned
earlier; see ref. \cite{DasPRE2025}), by removing the $1/T$ term to avoid its divergence
 in the $T = 0$ limit, as required for the quantum case here for the SK
model in transverse field (at $T = 0$).


The generalized mean field dynamics of the
Ising spins in presence of both longitudinal
and transverse field, extending the classical
Suzuki–Kubo formalism \cite{SuzukiKubo1968} can be represented
(cf. \cite{SuzukiKubo1968,DasPRE2025,TAP1977} and \cite{deGennes1963,Brout1966,SenChakrabarti1989,AcharyyaPhysicaA1994,AcharyyaJPhysA1994}) by the following
differential equation :

\begin{equation}
 \frac{d\vec{m}_i^{[c]}}{dt} = - \vec{m}_i^{[c]} +
\tanh\left(\frac{|\vec{h}_i^{\text{eff}\,[c]}|}{T}\right) \frac{\vec{h}_i^{\text{eff}\,[c]}}{\left|\vec{h}_i^{\text{eff}\,[c]}\right|} . \tag{6}
\end{equation}

\noindent The above vector differential equation is basically
first order nonlinear coupled differential equations
for $\vec{m}_i^{[c]} = \langle\vec{\sigma}_i^{[c]} \rangle$.
They can be explicitly rewritten as


\[
\frac{d m_i^{x\,[c]}}{dt} = - m_i^{x\,[c]} +
\tanh\left( \frac{ \left| \vec{h}_i^{\text{eff}\,[c]} \right| }{T} \right) \cdot \frac{\Gamma} { \left| \vec{h}_i^{\text{eff}\,[c]} \right| } \tag{7a}
\]

and


\[
\frac{d m_i^{z\,[c]}}{dt} = - m_i^{z\,[c]} +
\tanh\left( \frac{ \left| \vec{h}_i^{\text{eff}\,[c]} \right| }{T} \right) \cdot
\frac{ h_i^{z\,\text{eff}\,[c]}  }{ \left| \vec{h}_i^{\text{eff}\,[c]} \right| }.
\tag{7b}
\]

For discrete time ($t$), the above
equation can be simplified to:


\[
m_i^{x\,[c]}(t+1) =
\tanh\left(
\frac{ \left| \vec{h}_i^{\text{eff}\,[c]}(t) \right| }{T(t)}
\right)
\cdot
\frac{ \Gamma(t) }{ \left| \vec{h}_i^{\text{eff}\,[c]}(t) \right| }
\tag{8a}
\]

and

\[
m_i^{z\,[c]}(t+1) =
\tanh\left(
\frac{ \left| \vec{h}_i^{\text{eff}\,[c]} (t)\right| }{T(t)}
\right)
\cdot
\frac{ h_i^{z\,\text{eff}\,[c]} (t) }{ \left| \vec{h}_i^{\text{eff}\,[c]} (t)\right| },
\tag{8b}
\]
where $\left| \vec{h}_i^{\text{eff}\,[c]}(t) \right|$, $h_i^{z\,\text{eff}\,[c]}(t)$ are given by Eqs. (4a) and (4b) with the corresponding variables at time $t$. 
Note, the time
here (and also the continuous or discrete intervals) are  normalized  by the
microscopic relaxation time in Eq. (6).


\section{Numerical study}
We will solve numerically the above coupled equations (8) for $m_i^{z\,[c]}(t)$, using Eq. (3) and (4) for $\vec{h}_i^{\text{eff}\,[c]}(t)$, for an $N$ spin SK model ($N$ in the range 25 to 10,000) averaging over $[c]$ in the range 10000 to 15, respectively (the corresponding code is freely available in ref. \cite{code}). For the annealing dynamics of the system, the annealing parameters $T$ and $\Gamma$ start from the para phase ($q^{[c]}=0$) for $T(t=0)=T_0>T_c$ and/or $\Gamma(t=0)=\Gamma_0>\Gamma_c$ and following 
\begin{equation}
    T(t) = T(t=0)[1-t/\tau] \tag{9a}
 \end{equation}   
    \noindent and/or

\begin{equation}
    \Gamma(t) = \Gamma(t=0)[1-t/\tau]. \tag{9b} 
\end{equation}

\begin{figure}
\includegraphics[width=8.5cm]{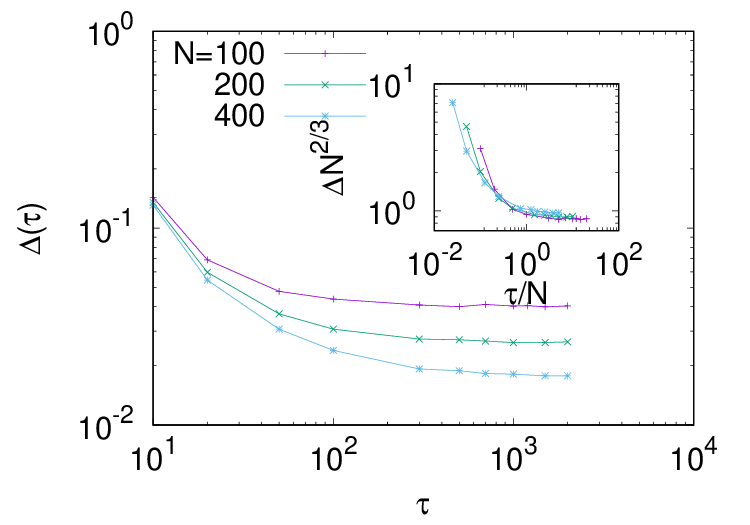}
\caption{The variations of $\Delta(\tau)=E_N^0-E^0$ are shown with annealing time ($\tau$) for quantum SK model (at $T=0$). Different system sizes are indicated. The time taken for the reaching the saturation in the energy difference scales linearly with the system size, as shown in the inset. The same order of annealing time is obtained for saturation in $\Delta$ for fixed $N$ in the classical \cite{DasPRE2025} and mixed annealing cases.}
\label{qa_tau}
\end{figure}

For the classical case ($\Gamma$ = 0 with $h=0$), this equation
(6a) for $dm_i^z/dt$ has been already studied 
for classical annealing in the SK model \cite{DasPRE2025}.
Here the annealing time $\tau$ refers
to the earliest time in which the configurational average over the
cooperative (longitudinal) spin-spin interaction energy part of the SK
model saturates to its ground state value (see Fig.  \ref{qa_tau}).

\subsection{Equilibrium phase diagram}

Before going for the annealing of the system, we demonstrate the phase boundary between the spin glass and para phase that could be obtained from solving the coupled equations (8) with the longitudinal field $h=0$. For these, the external parameters $T$ and $\Gamma$ are held fixed, and the system is allowed to relax, from an arbitrary initial condition, to the equilibrium state. The spin glass order parameter $q$ is measured (averaged over configurations) using Eq. (5), and that indicates the phase boundary between spin-glass and para phases. 

\begin{figure}[tbh]
\includegraphics[width=8.5cm]{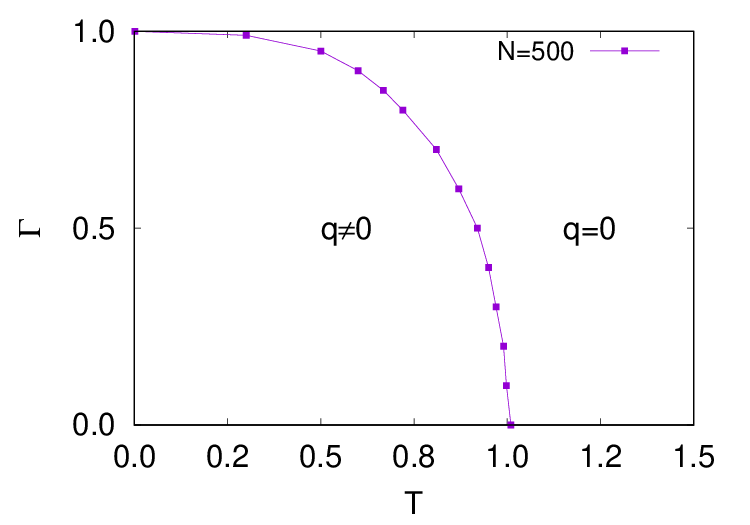}
\caption{The phase boundary for the SK model (cf. \cite{yamamoto}) is shown that separates the $q\ne0$ (spin glass ordered) and $q=0$ (para phase) for $N=500$.}
\label{qm_phase}
\end{figure}

In Fig. \ref{qm_phase}, the above mentioned phase boundary is shown for a system size $N=500$ with configuration average of 100.  The boundary suggests that the critical point for the temperature driven transition in the classical case ($\Gamma=0$) is given by $T_c=1$, as is for the classical mean-field SK model. On the other hand, the quantum transition ($T=0$) is obtained at $\Gamma_c=1$, which agrees with the thermofield dynamical estimate in ref. \cite{yamamoto}, but less than best estimates ($\Gamma_c(T=0) \approx 1.5$) \cite{sudip,cirac}.

The phase diagram suggests that although we have approximated the dynamics of the local magnetization through Suzuki-Kubo equation, the phase boundary remains almost the same. With this, we can now go on to study the annealing dynamics to reach the ground state of the model under classical, quantum and intermediate regimes.

\subsection{Classical annealing}
\begin{figure}
\includegraphics[width=8.5cm]{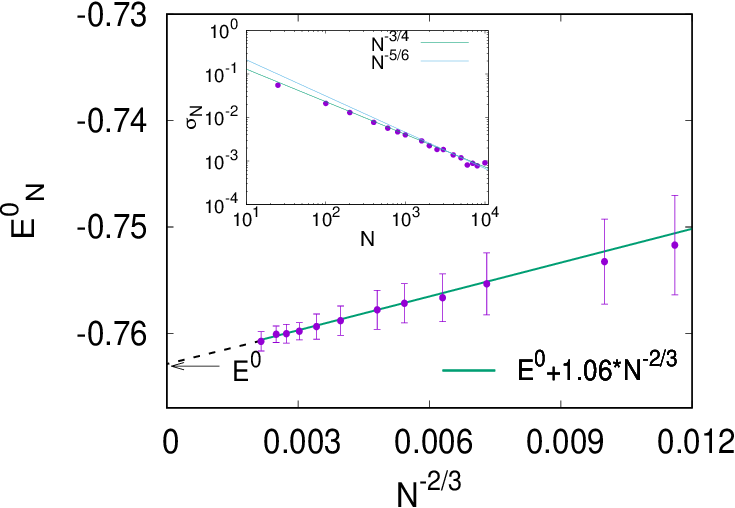}
\caption{Classical annealing with time variation of $T$ given by Eq. (9a) (with $\Gamma=0$, $h=0$): The lowest energy values for given system size are plotted against $N^{-2/3}$ which shows a scaling $E_{0}(N)\sim N^{-2/3}$. The ground state energy per spin ($N \rightarrow {\infty}$) is $E^0=-0.763166 \dots$. From the least-square fitting we get a ground state energy which is $E^0=-0.7629\pm 0.0002$ (considering the exponent to be 2/3). The inset shows the variation of the fluctuations $\sigma_N$ of $E^0_N$ with system size ($\sigma_N\sim N^{-3/4}$) and $\sigma_N\sim N^{-5/6}$).}
\label{qm_en_c}
\end{figure}
The classical annealing of the model implies $\Gamma=0$ and the temperature varying following Eq. (9(a)). These results have already been reported elsewhere \cite{DasPRE2025}, but we are including here for the completeness of the discussion.  

Fig. \ref{qm_en_c} shows the estimate of the ground state energy $E^0$ per spin, at the end of the annealing process, is plotted for different system sizes, showing a variation of the form $E^0_N=E^0+aN^{-2/3}$, where $E^0 = -0.7629 \pm 0.0002$, which is close to the best numerical estimate ground state energy for the thermodynamic limit \cite{boet,pala} and $a$ is a constant. The fluctuation in the ground state energy estimate $\sigma_N$ decays with system size as $N^{-5/6}$, which is shown in the inset. Both of these exponent values are well matched with known estimates.  This annealing approach is very simple and algorithmically affordable ($N^3$). 

\subsection{Quantum annealing}

\begin{figure}
\includegraphics[width=8.5cm]{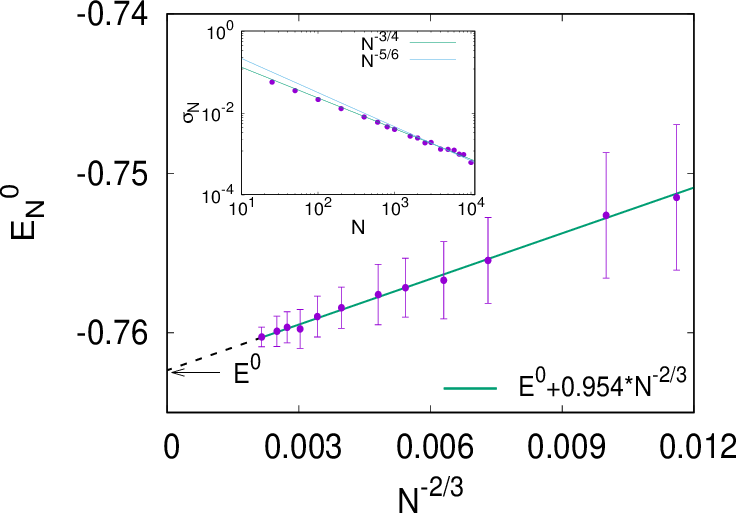}
\caption{Quantum annealing with time variation of transverse field $\Gamma$ following Eq. (9b) ($T=0$, $h=0$): The lowest energy values for given system size are plotted against $N^{-2/3}$ which shows a scaling $E_{0}(N)\sim N^{-2/3}$ ($N$ in the range 25 to 10000). The ground state energy per spin ($N \rightarrow {\infty}$) is $E^0=-0.763166 \dots$. From the least-square fitting we get a ground state energy which is $E^0=-0.7623\pm 0.0001$ (considering the exponent to be 2/3). The inset shows the variations of the fluctuations $\sigma_N$ of $E^0_N$ with system size ($\sigma_N\sim N^{-3/4}$ and $\sigma_N\sim N^{-5/6}$).}
\label{qm_en_q}
\end{figure}

The simplicity and affordability of the classical annealing using Suzuki-Kubo dynamics naturally raises interests in the quantum equivalent of the approach. For a purely quantum annealing, the temperature is set to zero and the transverse field is varied during the dynamics, following Eq. (9(b)).

In Fig. \ref{qm_en_q}, the estimates of the ground state energies are plotted for various system sizes, as before. It turns out, the estimated values for each of the system size, are close to what was found for purely classical annealing. Indeed, the power law variations with the system size remain the same for both the ground state energies and its fluctuations. The annealing time $\tau$, used for the dynamics also has the same linear scaling as in the classical case. This implies that the algorithmic cost of the simulation remains $N^3$. The extrapolated value of the ground state energy per spin ($E^0=-0.7623\pm 0.0001$) in the infinite system size limit is also very close to the classical case ($E^0=-0.7629\pm 0.0002$).  These results indicate, therefore, that the algorithmic advantage here comes from the fact that the magnetization variables are made continuous due to the Suzuki-Kubo dynamics. The case of quantum annealing, which effectively make the magnetization variable continuous through both longitudinal and transverse components, does not bring any added advantage.  

\subsection{Mixed annealing}

\begin{figure}
\includegraphics[width=8.5cm]{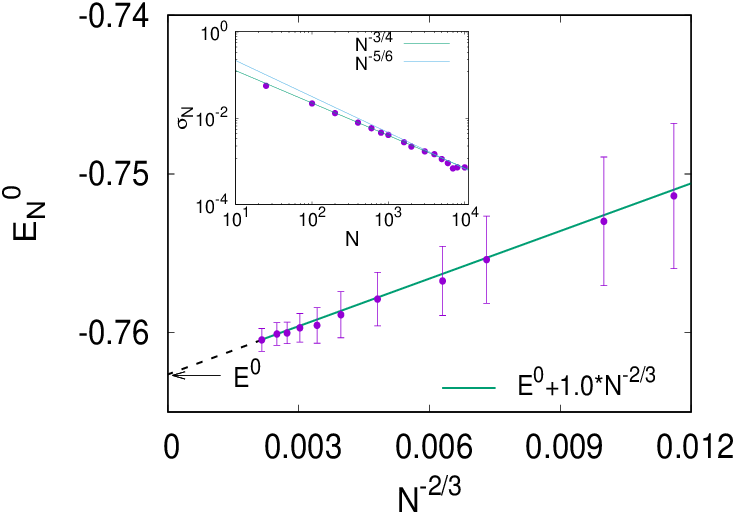}
\caption{Mixed annealing (following Eqs. (9a,9b), $h=0$)  starting from the critical phase boundary ($T_c=0.94$,$\Gamma_c=0.5$): The lowest energy values for given system size are plotted against $N^{-2/3}$ which shows a scaling $E_{0}(N)\sim N^{-2/3}$ ($N$ in the range 25 to 10000). The ground state energy per spin ($N \rightarrow {\infty}$) is $E^0=-0.763166 \dots$. From the least-square fitting we get a ground state energy per spin which is $E^0=-0.7626\pm 0.0001$ (considering the exponent to be 2/3). The inset shows the variations of the fluctuations $\sigma_N$ of $E^0_N$ with system size ($\sigma_N\sim N^{-3/4}$ and $\sigma_N\sim N^{-5/6}$).}
\label{qm_en_cq}
\end{figure}

Finally we look into the case where both the temperature and transverse fields are varied simultaneously during the annealing process following Eqs. (9(a)) and (9(b)). The time scale for annealing, $\tau$, is kept the same for both of these parameters. In Fig. \ref{qm_en_cq}, the ground state energy estimates and its fluctuations are plotted. The starting point of the annealing process was at $T_c= 0.94$ and $\Gamma_c=0.5$. Once again, the energy values are very close to what was obtained for purely classical and quantum cases, with the same power law variations, albeit with a slightly different value for the constant $a$. The extrapolated value of the ground state energy in the thermodynamic limit is $E^0=-0.7626\pm 0.0001$. Of course, the algorithmic cost also remains the same ($N^3$). 

\begin{figure}
\includegraphics[width=8.5cm]{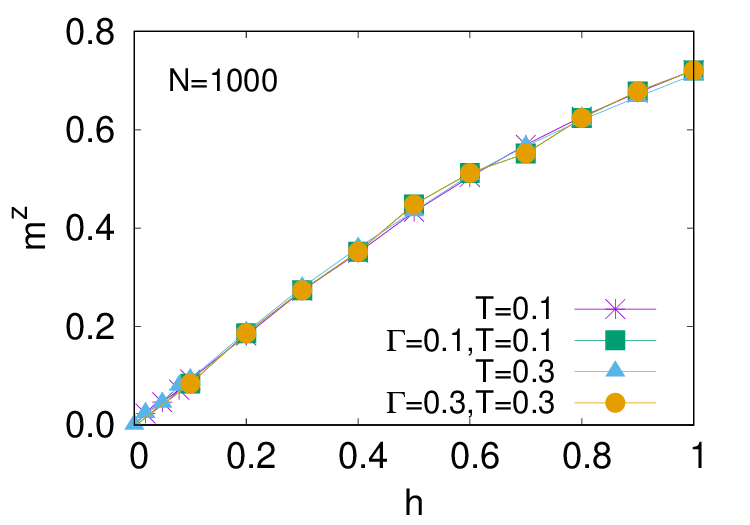}
\caption{The $z$-component of the magnetization is plotted against the longitudinal field $h$ for different temperatures, which suggests that there is no AT line for this dynamical Eq. (7) or (8). The errors are smaller than the symbol sizes.}
\label{mz_h}
\end{figure}

\begin{table}
\caption{Quantum Annealing: The number of configuration averages done for each
system size ($N$) and the corresponding estimates for the
ground state energies ($E^0_N$)  and their errors (estimated from
the standard deviations) are listed below.}
    \centering
    \setlength{\tabcolsep}{5.0pt} 
\renewcommand{\arraystretch}{1.0}
    \begin{tabular}{|l|c|r|}
    \hline
        System size ($N$) & Configs. & Ground state energy ($E^0_{N}$) \\ 
        \hline
$25$ &$10000$  & $-0.669{\pm}0.056$ \\  \hline
$100$ & $5000$ &  $-0.722{\pm}0.021$\\ \hline
$200$ & $5000$ & $-0.736{\pm}0.013$ \\ \hline
$400$ & $2000$ & $-0.745{\pm}0.008$ \\ \hline
$600$ & $1000$ & $-0.749{\pm}0.006$ \\ \hline
$800$ & $800$ & $-0.751{\pm}0.004$ \\ \hline
$1000$ & $500$ &  $-0.753{\pm}0.004$\\ \hline
$1600$ & $300$ & $-0.755{\pm}0.003$ \\ \hline
$2000$ & $100$ & $-0.757{\pm}0.002$ \\ \hline
$3000$ & $80$ & $-0.758{\pm}0.002$  \\ \hline
$4000$ & $50$ & $-0.758{\pm}0.001$ \\ \hline
$5000$ & $50$ & $-0.759{\pm}0.001$ \\ \hline
$6000$ & $30$  & $-0.760{\pm}0.001$ \\ \hline
$7000$ & $50$  & $-0.760{\pm}0.001$ \\ \hline
$8000$ & $20$  & $-0.760{\pm}0.001$ \\ \hline
$10000$ & $24$ & $-0.7602{\pm}0.0006$ \\ \hline
    
    \end{tabular}
    \label{qm_tab}
\end{table}

\begin{table}
\caption{Mixed Annealing: The number of configuration averages done for each
system size ($N$) and the corresponding estimates for the
ground state energies ($E^0_N$)  and their errors (estimated from
the standard deviations) are listed below.}
    \centering
    \setlength{\tabcolsep}{5.0pt} 
\renewcommand{\arraystretch}{1.0}
    \begin{tabular}{|l|c|r|}
    \hline
        System size ($N$) & Configs. & Ground state energy ($E^0_{N}$) \\ 
        \hline
$25$ &$10000$  & $-0.668{\pm}0.055$ \\  \hline
$100$ & $5000$ &  $-0.723{\pm}0.021$\\ \hline
$200$ & $5000$ & $-0.737{\pm}0.013$ \\ \hline
$400$ & $2000$ & $-0.746{\pm}0.008$ \\ \hline
$600$ & $1000$ & $-0.750{\pm}0.006$ \\ \hline
$800$ & $800$ & $-0.751{\pm}0.004$ \\ \hline
$1000$ & $400$ &  $-0.753{\pm}0.004$\\ \hline
$1600$ & $300$ & $-0.755{\pm}0.003$ \\ \hline
$2000$ & $200$ & $-0.757{\pm}0.002$ \\ \hline
$3000$ & $80$ & $-0.758{\pm}0.002$  \\ \hline
$4000$ & $50$ & $-0.759{\pm}0.001$ \\ \hline
$5000$ & $50$ & $-0.759{\pm}0.001$ \\ \hline
$6000$ & $30$  & $-0.760{\pm}0.001$ \\ \hline
$7000$ & $30$  & $-0.760{\pm}0.001$ \\ \hline
$8000$ & $20$  & $-0.760{\pm}0.001$ \\ \hline
$10000$ & $15$ & $-0.7604{\pm}0.0007$ \\ \hline
    
    \end{tabular}
    \label{mix_tab}
\end{table}

\section{Discussion and Conclusion}
Here we have studied the quantum annealing dynamics of the Sherrington-Kirpatrik mean field spin glass. We have used the Suzuki-Kubo mean field dynamical equations for the individual local magnetization variables, which makes the individual magnetization variables continuous in $(-1,1)$, and is augmented by de Genne's mean field quantum dynamics \cite{deGennes1963,Brout1966}.  The effective field on the individual spins then results from the Hamiltonian and the modified TAP reaction field (to fit the $T=0$ quantum case; see \cite{DasPRE2025}). The dynamical evolution of the individual components, both transverse and longitudinal, of the magnetization variables could then be written as coupled differential equations, which we then discretize and solve numerically. 

\begin{figure}[tbh]
\includegraphics[width=8.5cm]{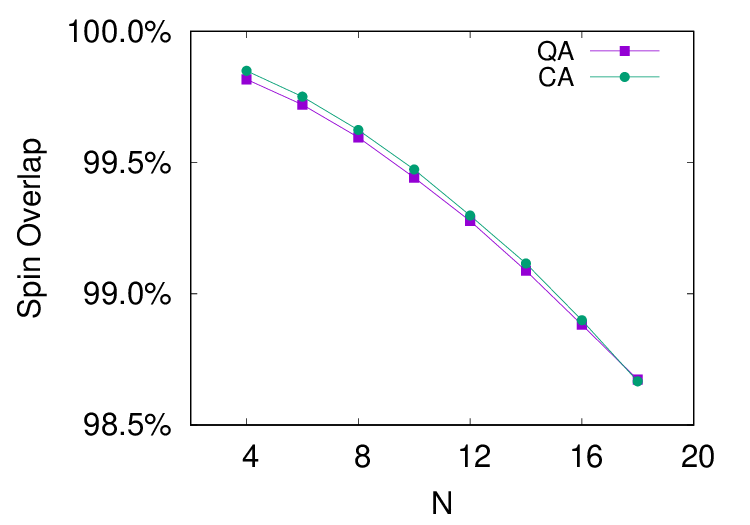}
\caption{The exact percentage of spin overlap between the ground state obtained from quantum annealing (QA) and classical annealing (CA) through the Suzuki-Kubo dynamics and the exact ground state obtained from brute force method. The errors are smaller than the symbol sizes.}
\label{hamm_dist}
\end{figure}
This method of approximation has been applied before for the classical annealing and that resulted in a fast algorithm (algorithmic cost of $N^3$) in estimating the ground state energy. Here we have done that for the quantum annealing (keeping $T=0$), and a mixed annealing, where both $T$ and $\Gamma$ were varied in the same time scale. 

The estimates of the ground state energies, both for the largest system size ($N=10000$) simulated here and for the extrapolated value, remain similar for all three cases: Classical ($E^0 =-0.7629 \pm 0.0002$), Quantum ($E^0=-0.7623 \pm 0.0001$) and Mixed ($E^0=-0.7626 \pm 0.0001$), all of which are to be compared with the best known estimate $E^0= -0.763166726 \dots$ \cite{Oppermann2007}. It is interesting to note that the introduction of the transverse field, which otherwise helps reach the ground state much faster, does not significantly influence either the energy estimates or the algorithmic cost.  The probable reason for this is that in the Suzuki-Kubo approximation, the $m_i^z$ are already continuous making the corrugated free energy landscape to be approximately flat towards the beginning of the annealing dynamics, allowing the configuration to escape any potential trapped state. At the end, when $T\to 0$, the energy landscape regains its structure, but the configuration already resides near the ground state. When the transverse field is introduced, the spin will have two components now $m_i^x$ and $m_i^z$, which again allows $m_i^z$ to be continuous. But given this was already the case for the classical case (even without $m_i^x$), it does not bring any additional gain. Similarly, the estimates are close for the mixed annealing case as well. 

Note that in all these cases, we have taken the starting point of the annealing at the $\Gamma-T$ phase boundary. If the starting points are away from the phase boundary, the final estimate of the ground state energy becomes higher.

Although the estimates of the ground state energies are quite good under this approximation, we have checked for the de Almeida-Thouless (AT) line when a longitudinal field is applied \cite{AT,BinderYoung1986}. We do not see such effect here both for the classical case as well as for the quantum case. In Fig. \ref{mz_h}, the average magnetization is plotted against the longitudinal field (with constant $T$ and $\Gamma$), which shows a linear growth even for $h\to 0^+$ limit. The absence of the AT line is also perhaps due to the continuous nature of the magnetization components in the Kubo-Suzuki-de Gennes quantum mean field equations (7-8) used here. For quantum case, these continuous magnetization components are built-in and not any approximation, as in the classical case. For the quantum SK model case this absense of AT-line has already been observed and reported \cite{cirac}.  

Finally, we have also checked the overlaps of the annealed states and the actual ground states. We could find the actual ground states by brute force for small system sizes. The annealing methods described here was then applied for such systems. We then calculate the overlaps between the final configuration after annealing and the actual ground states for a given set of $J_{ij}$ values and then average over different sets. Fig. \ref{hamm_dist} shows the results for system sizes upto $N=18$ in both classical and quantum cases. The high overlaps suggest that the estimates obtained here are fairly good for the ground states. 

In conclusion, we have developed a quantum annealing approach utilizing the Suzuki-Kubo dynamical equations for the SK spin glass model. The method gives a fast algorithm to make a reasonably good estimate of the ground state and energy. The estimates for classical, quantum and mixed annealing approaches employing Suzuki-Kubo mean field dynamics give same values (within error bars $\sigma_N$ varying with system size $N$ as $\sigma_N\sim N^{-5/6}$) for
the ground state energy per spin $E^0$ of the model (converging to its best estimates \cite{Oppermann2007}; scaling as $N^{-2/3}$ with system size $N$) and also within the same order of annealing time
$\tau$ (linear in $N$). This indicates that there is no ``quantum supremacy" (cf.
\cite{suz}),  when one employs the Suzuki-Kubo mean field dynamics for annealing of the de Gennes version of the quantum Ising Sherrington-Kirkpatrick model.

\section*{Data availability statement}
The data used in the manuscript were generated from a code, which is freely available at https://github.com/soumya-
84/SK\_Quantum/blob/main/qm\_annealing\_ga1.c .

\section*{Author contributions}
Soumyaditya Das: Investigation, Visualization, Numerical simulation, writing; Soumyajyoti Biswas: Conceptualization, analysis, writing; Bikas K. Chakrabarti: Conceptualization, analysis, writing.

\section*{Acknowledgements} We are thankful to Muktish Acharyya for useful discussions.
Bikas K. Chakrabarti is grateful to the Indian National Science Academy for their
Senior Scientist Research Grant. The simulations were performed using  HPCC Chandrama in SRM University-AP.

\end{document}